\documentclass[copyright,creativecommons]{eptcs}
\providecommand{\event}{QAPL 2010} 
\providecommand{\volume}{??}
\providecommand{\anno}{2010}
\providecommand{\firstpage}{1}
\providecommand{\eid}{??}

\usepackage{amsmath}
\usepackage{amsfonts}
\usepackage{amssymb}
\usepackage{stmaryrd}
\usepackage{verbatim}
\usepackage{pst-all}
\usepackage[dvips,xdvi]{graphicx}
\usepackage{graphics}

\usepackage{breakurl}        

\usepackage{stmaryrd}
\usepackage{verbatim}
\usepackage{pslatex}
\usepackage{floatflt}

\newtheorem{theorem}{Theorem}[section]
\newtheorem{lemma}[theorem]{Lemma}

\newtheorem{corollary}[theorem]{Corollary}
\newenvironment{proof}[1][Proof]{\begin{trivlist}
\item[\hskip \labelsep {\bfseries #1}]}{\end{trivlist}}
\newenvironment{proofsketch}[1][Proof sketch]{\begin{trivlist}
\item[\hskip \labelsep {\bfseries #1}]}{\end{trivlist}}
\newcommand{\qed}{\hfill $\square$ }

\title{Optimal Time-Abstract Schedulers for CTMDPs\\ and Markov Games%
\thanks{This work was partly supported by the German Research
Foundation (DFG) as part of the Transregional Collaborative Research
Center ``Automatic Verification and Analysis of Complex Systems''
(SFB/TR 14 AVACS) and by the Engineering and Physical Science Research Council (EPSRC) through grant EP/H046623/1 ``Synthesis and Verification in Markov Game Structures''.}
}
\author{Markus Rabe
\institute{
Universit\"at des Saarlandes}
\email{rabe@cs.uni-saarland.de}
\and Sven Schewe
\institute{
University of Liverpool}
\email{sven.schewe@liverpool.ac.uk}
}
\def\titlerunning{Optimal Time-Abstract Schedulers for CTMDPs and Markov Games}
\def\authorrunning{Markus Rabe and Sven Schewe}

\newcommand{\set}[1]{ \left\{ #1 \right\} } 
\newcommand{\R}{\mathbb{R}} 
\newcommand{\N}{\mathbb{N}} 
\newcommand{\Q}{\mathbb{Q}} 

\newcommand{\locations}{{L}}
\newcommand{\act}{\mathit{Act}} 
\newcommand{\ratematrix}{\mathbf{R}}
\newcommand{\probabilitymatrix}{\mathbf{P}}

\newcommand{\nM}{\ensuremath{n_{\mathcal M}}}
\newcommand{\M}{{\ensuremath\mathcal M }}
\newcommand{\U}{{\ensuremath\mathcal U }}
\renewcommand{\S}{\ensuremath{\mathcal{S}}}
\newcommand{\G}{\ensuremath{\mathcal{G}}}

\newcommand{\dist}{\mathit{Dist}}
\newcommand{\paths}{\mathit{Paths}}
\newcommand{\pathsabs}{\mathit{Paths}_{\mathit{abs}}}

\newcommand{\prob}{\mathit{Pr}}
\newcommand{\Prob}{\mathit{PR}}

\newcommand\shift{\mathit{shift}}

\newcommand\vis{\mathit{vis}}

\begin{document}

\maketitle

\begin{abstract}

We study time-bounded reachability in continuous-time Markov decision processes for time-abstract scheduler classes.
Such reachability problems play a paramount r\^ole in dependability analysis and the modelling of manufacturing and queueing systems.
Consequently, their analysis has been studied intensively, and techniques for the approximation of optimal control are well understood.
From a mathematical point of view, however, the question of approximation is secondary compared to the fundamental question whether or not optimal control exists.

We demonstrate the existence of optimal schedulers for the time-abstract scheduler classes for all CTMDPs. Our proof is constructive: We show how to compute optimal time-abstract strategies with finite memory. 
It turns out that these optimal schedulers have an amazingly simple structure---they converge to an easy-to-compute memoryless scheduling policy after a finite number of steps. 

Finally, we show that our argument can easily be lifted to Markov games:
We show that both players have a likewise simple optimal strategy in these more general structures.
\end{abstract}

\section{Introduction}

Markov decision processes (MDPs) are a framework that incorporates both nondeterministic and probabilistic choices.
They are used in a variety of applications such as the control of manufacturing processes \cite{Puterman1994,Feinberg/04/MDP} or queueing systems \cite{Sennott/99/queueing}.
We study a real time version of MDPs, continuous-time Markov decision processes (CTMDPs), which are a natural formalism for modelling in scheduling \cite{BDF/81/scheduling,Puterman1994} and stochastic control theory \cite{Feinberg/04/MDP}.
CTMDPs can also be seen as a unified framework for different stochastic model types used in dependability analysis \cite{sanders1989rbm,Puterman1994,GeneralizedStochasticPetriNetsAjmone95,hermanns2002imc,NeuhausserDelayedNondeterminism09}.

The analysis of CTMDPs usually concerns the different possibilities to resolve the nondeterminism by means of a scheduler (also called strategy). Typical questions cover qualitative as well as quantitative properties, such as: ``Can the nondeterminism be resolved by a scheduler such that a predefined property holds?'' or respectively ``Which scheduler optimises a given objective function?''.

As a slight restriction, nondeterminism is either always hostile or always supportive in CTMDPs.
Markov games \cite{CompetitiveMarkovDecisionProcesses} provide a generalisation of CTMDPs by disintegrating the control locations into locations where the nondeterminism is resolved angelically (supportive nondeterminism) and control locations where the nondeterminism is resolved demonically (hostile nondeterminism).

In this paper, we study the \emph{maximal time-bounded reachability problem} \cite{Puterman1994,Baier+all/05/efficientCTMDP,ZHHW08,NeuhausserDelayedNondeterminism09,neuhausserzhangTimeBoundedReachabilityReport,brazdil_et_al:LIPIcs:2009:2307} in CTMDPs and Markov games. 
Time-bounded reachability is the standard control problem to construct a scheduler that controls the Markov decision process such that the likelihood of reaching a goal region within a given time bound is maximised, and to determine the probability. 
For games, both the angelic and the demonic nondeterminism needs to be resolved at the same time.

The obtainable quality of the resulting scheduling policy naturally depends on the power a scheduler has to observe the run of the system and on its ability to store and process this information.
The commonly considered schedulers classes and their basic connections have been discussed in the literature~\cite{NeuhausserDelayedNondeterminism09, WolovickJohr06Meaningful}. 
Thereof, we consider those schedulers that have no direct access to time, the time-abstract schedulers. 
The time-abstract scheduler classes that can observe the history, its length, or nothing at all, are marked H (for history-dependent), C (for hop-counting), and P (for positional), respectively.

These classes form a simple inclusion hierarchy ($\text{H} \supset \text{C} \supset \text{P}$) and in general they yield different maximum reachability probabilities. However, it is known that for uniform CTMDPs the maximum reachability probabilities of classes H and C coincide~\cite{Baier+all/05/efficientCTMDP}. 
Uniform CTMDPs have a uniform transition rate $\lambda$ for all their actions. 

\vspace{-.1cm}
\paragraph{Optimal schedulers.} Given its practical importance, the bounded reachability problem for Markov decision processes (and their deterministic counterpart the \emph{Markov chains}) has been intensively studied~\cite{Aziz/00/exactModelCheckingCTMC, Baier+all/05/efficientCTMDP, NeuhausserDelayedNondeterminism09, brazdil_et_al:LIPIcs:2009:2307}. 


While previous research focused on \emph{approximating} optimal scheduling policies~\cite{Baier+all/05/efficientCTMDP}, the existence of optimal schedulers for all scheduler classes has been demonstrated in Rabe's master thesis \cite{Rabe:Thesis:2009,atr55}, on which this paper is partly based.
Meanwhile, Brazdil et al.~\cite{brazdil_et_al:LIPIcs:2009:2307} have independently provided a similar result for \emph{uniform} Markov games, that is, for games that use the same transition rate for all actions.



\vspace{-.1cm}
\paragraph{\bf Contribution.}
We start with a report on our work on counting (C) and history dependent (H) schedulers in \emph{uniform} CTMDPs. 
Although the case of the  counting schedulers could by now be inferred as a corollary from the existence of optimal counting strategies in Markov games~\cite{brazdil_et_al:LIPIcs:2009:2307}, we decided to present it for 2.5 reasons: 
Firstly, it requires only marginal extra effort. 
Secondly, CTMDPs have been an important object of study for decades whereas Markov games are comparably new, and we think that our proof can provide insights in particular to readers that are not familiar with games. 
Finally, it was developed independently and at the same time. 



We then show how our result on uniform CTMDPs can be lifted to general CTMDPs, and that randomisation cannot improve the quality of optimal scheduling. 
In Section~\ref{sect:games}, we show that our lifting argument naturally extends to Markov games:
We show that there are optimal time-abstract counting and history dependent schedulers with finite memory for general Markov games and that---as for CTMDPs---randomisation cannot improve optimal scheduling for either player. 

Our solution builds on the observation that, if time has almost run out, we can use a greedy strategy that optimises our chances to reach our goal in fewer steps rather than in more steps. 
We show that a memoryless greedy scheduler exists, and is indeed optimal after a certain step bound. 
The existence of an optimal scheduler is then implied by the finite number of remaining candidates---it suffices to search among those schedulers that deviate from the greedy strategy only in a finite preamble.

The extension to non-uniform CTMDPs (and Markov games) builds upon a simple uniformisation technique and draws from a class of schedulers that are (partially) blind to the additional information introduced by the uniformisation. 
%
With the help of this scheduler class, we successively demonstrate that it is optimal (in the game case for both players) to turn to a fixed memoryless greedy strategy after a finite number of steps that is easy to compute.
Hence, we can focus on scheduling policies that deviate from this scheduling policy only on a finite preamble.
It then suffices to exclude that randomisation can improve the result (for either player) to reduce the candidate strategies to a finite set, and hence to infer the existence of simple optimal strategies for the non-uniform case as well.

\section{Continuous-Time Markov Decision Processes}\label{sect:definitions}

A \emph{continuous-time Markov decision process} $\M$ is a tuple $(\locations,\act,\ratematrix,\nu,B)$ with a finite set of locations $\locations$, a finite set of actions $\act$, a rate matrix $\ratematrix: (\locations\times\act\times\locations) \to \Q_{\geqslant0}$, an initial distribution $\nu\in\dist(\locations)$, and a goal region $B\subseteq\locations$. 
We define the total exit rate for a location $l$ and an action $a$ as $\ratematrix(l,a,\locations) = \sum_{l'\in\locations}{\ratematrix(l,a,l')}$. 
For a CTMDP we require that, for all locations $l\in\locations$, there must be an action $a\in\act$ such that $\ratematrix(l,a,\locations)>0$, and we call such actions \emph{enabled}. 
We define $\act(l)$ to be the set of enabled actions in location $l$.
If there is only one enabled action per location, a CTMDP $\M$ is a continuous-time Markov chain \cite{280952}. If multiple actions are available, we need to resolve the nondeterminism by means of a scheduler (also called strategy or scheduling policy). As usual, we assume the goal region to be absorbing, and we use $\probabilitymatrix(l,a,l')=\frac{\ratematrix(l,a,l')}{\ratematrix(l,a,\locations)}$ to denote the time-abstract transition probability. 


\paragraph{\bf Uniform CTMDPs. } We call a CTMDP uniform with rate $\lambda$ if, for every location $l$ and action $a\in\act(l)$, the total exit rate $\ratematrix(l,a,\locations)$ is $\lambda$. In this case the probability $p_{\lambda t}(n)$ that there are exactly $n$ discrete events (transitions) in time $t$ is Poisson distributed: $p_{\lambda t}(n)=e^{-\lambda \, t}\cdot \frac{(\lambda\, t)^n}{n!}$.

We define the \emph{uniformisation} $\U$ of a CTMDP $\M$ as the uniform CTMDP obtained by the following transformation steps. 
We create a copy $l_{\U}$ for every $l\in\locations$ and obtain $\locations_\U=\bigcup_{l\in\locations}\set{l,l_\U}$. 
We call the new copies unobservable, and all locations $l\in\locations$ observable.
Let $\lambda$ be the maximal total exit rate in $\M$.
The new rate matrix $\ratematrix_{\U}$ extends $\ratematrix$ by first adding the rate $\ratematrix_{\U}(l,a,l_{\U})=\lambda-\ratematrix(l,a,\locations)$ for every location $l\in\locations$ and action $a\in\act$ of $\M$, and by then copying the outgoing transitions from every observable location $l$ to its unobservable counterpart $l_{\U}$, while the other components remain untouched. The intuition behind this uniformisation technique is that it enables us to distinguish whether a step would have occurred in the original automaton or not. 


\paragraph{\bf Paths.} A \emph{timed path} in CTMDP $\M$ is a finite sequence in $(\locations\times\act\times\R_{\geqslant0})^*\times\locations=\paths(\M)$.
We write
\[
  l_0\xrightarrow{a_0,t_0} l_1\xrightarrow{a_1,t_1}
    \cdots ~ \xrightarrow{a_{n-1},t_{n-1}}l_n
\]
for a sequence $\pi,$ and we require $t_{i-1}<t_{i}$ for all $i<n$. The $t_i$ denote the system's time when the events happen.
The corresponding \emph{time-abstract path} is defined as $l_0\xrightarrow{a_0} l_1\xrightarrow{a_1} \cdots ~ \xrightarrow{a_{n-1}}l_n$.
We use $\pathsabs(\M)$ to denote the set of all such projections and $\mid\cdot\mid$ to count the number of actions in a path.
Concatenation of paths $\pi,\pi'$ will be written as $\pi\circ\pi'$ if the last location of $\pi$ is the first location of $\pi'$.


\paragraph{\bf Schedulers.}

The system's behaviour is not fully determined by the CTMDP, we additionally need a scheduler that resolves the nondeterminism that occurs in locations where multiple actions are enabled. 
When analysing properties of a CTMDP, such as the reachability probability, we usually quantify over a class of schedulers. 
In this paper, we consider the following common scheduler classes, which differ in their power to observe and distinguish events:
%
%
\begin{itemize}
	\item[$\circ$] \emph{Time-abstract history-dependent} (H) schedulers \hfill
	   $\pathsabs(\M)\rightarrow D$ \hspace{1cm}\mbox{}\\
	   that map time-abstract paths to decisions. 
	\item[$\circ$] \emph{Time-abstract hop-counting} (C) schedulers \hfill
	   $\locations\times\N\rightarrow D$ \hspace{1cm}\mbox{}\\
	   that map locations and the  length of the path to decisions. 
	\item[$\circ$] \emph{Positional} (P) or memoryless schedulers \hfill
	   $\locations\rightarrow D$ \hspace{1cm}\mbox{}\\
	   that map locations to decisions. 
\end{itemize}
Decisions $D$ are either randomised (R), in which case $D = \dist(\act)$ is the set of distributions over enabled actions, or are restricted to deterministic (D) choices, that is $D = \act$.
Where it is necessary to distinguish randomised and deterministic versions we will add a postfix to the scheduler class, for example HD and HR. 
We restrict all scheduler classes to those schedulers creating a measurable probability space~(cf.\ \cite{WolovickJohr06Meaningful}).


\paragraph{\bf Induced Probability Space. }

We build our probability space in the natural way: we first define the probability measure for cylindric sets of paths that start with
\[  l_0\xrightarrow{a_0,t_0} l_1\xrightarrow{a_1,t_1} 
    \cdots ~ \xrightarrow{a_{n-1},t_{n-1}}l_n,
\]
with $t_j \in I_j$ for all $j<n$, and for non-overlapping open intervals $I_0, I_1, \ldots, I_{n-1}$, to be the usual probability that a path starts with these actions for a given randomised scheduler $\S$
, and such that $\S(l_0\xrightarrow{a_{0},t_{0}}\dots\xrightarrow{a_{i-1},t_{i-1}} l_i)$ is equivalent for all $(t_0,\ldots,t_{i-1})\in I_0 \times \ldots \times I_{i-1}$:
\[
\int_{t_0 \in I_0, t_1 \in I_1,\dots, t_{n-1} \in I_{n-1}}\prod_{i=0}^{n-1}
\S(l_0\xrightarrow{a_{0},t_{0}}\dots\xrightarrow{a_{i-1},t_{i-1}} l_i)(a_i)\cdot
\ratematrix(l_i,a_i,l_{i+1}) \cdot
 e^{-\ratematrix(l_i,a_i,\locations)(t_i-t_{i-1})},
\]
assuming $t_{-1}=0$.

From this basic building block, we build our probability measure for measurable sets of paths and measurable schedulers in the usual way (cf.~\cite{WolovickJohr06Meaningful}).

\paragraph{\bf Time-Bounded Reachability Probability. }
For a given CTMDP $\M=(\locations,\act,\ratematrix,\nu,B)$ and a given measurable scheduler $\S$ that resolves the nondeterminism, we use the following notations for the probabilities:
\begin{itemize}
\item[$\circ$] $\prob_{\S}^{\M}(l,t)$ is the probability of reaching the goal region $B$ in time $t$ when starting in location $l$,

\item[$\circ$] $\prob_{\S}^{\M}(t) = \sum_{l \in \locations}\nu(l)\prob_{\S}^{\M}(l,t)$ denotes the probability of reaching the goal region $B$ in time $t$,

\item[$\circ$] $\prob_{\S}^{\M}(t;k)$ denotes the probability of reaching the goal region $B$ in time $t$ \emph{and} in at most $k$ discrete steps, and

\item[$\circ$] $\Prob_{\S}^{\M}(\pi,t)$ is the probability to traverse the time-abstract path $\pi$ 
within time $t$.
\end{itemize}

As usual, the supremum 
of the time-bounded reachability probability over a particular scheduler class is called the time-bounded reachability of $\M$ for this scheduler class, and we use `$\max$' instead of `$\sup$' to indicate that this value is taken for some \emph{optimal scheduler} $\S$ of this class.

\paragraph{\bf Step Probability Vector. }
Given a scheduler $\S$ and a location $l$ for a CTMDP $\M$, we define the \emph{step probability vector} $d_{l,\S}$ of infinite dimension. An entry $d_{l,\S}[i]$ for $i\geq0$ denotes the probability to reach goal region $B$ in up to $i$ steps from location $l$ (not considering any time constraints).

\section{Optimal Time-Abstract Schedulers}\label{sec:tabs}

In this section, we show that \emph{optimal} schedulers exist for all natural time-abstract classes, that is, for CD, CR, HD, and HR.
Moreover, we show that there are optimal schedulers that become positional after a small number of steps, which we compute with a simple algorithm.
We also show that randomisation does not yield any advantage: deterministic schedulers are as good as randomised ones.
Our proofs are constructive, and thus allow for the construction of optimal schedulers.
This also provides the first procedure to precisely determine the time-bounded reachability probability, because we can now reduce this problem to solving the time-bounded reachability problem of continuous-time Markov chains~\cite{Aziz/00/exactModelCheckingCTMC}.

Our proof consists of two parts. We first consider the class of uniform CTMDPs, which are much simpler to treat in the time-abstract case, because we can use Poisson distributions to describe the number of steps taken within a given time bound.
For uniform CTMDPs it is already known that the supremum over the bounded reachability collapses for all time-abstract scheduler classes from CD to HR \cite{Baier+all/05/efficientCTMDP}.
It therefore suffices to show that there is a CD scheduler 
which takes this value.

We then show that a similar claim holds for CD and HD scheduler in the general class of not necessarily uniform CTMDPs.
In this case, it also holds that there are simple optimal schedulers that converge against a positional scheduler after a finite number of steps, and that randomisation does not improve the time-bounded reachability probability.
However, in the non-uniform case the time-abstract path contains more information about the remaining time than its length only, and bounded reachability of history-dependent and counting schedulers usually deviate (see \cite{Baier+all/05/efficientCTMDP} for a simple example).

We start this section with the introduction of \emph{greedy schedulers}, HD schedulers that favour reachability in a small number of steps over reachability with a larger number of steps;
the positional schedulers against which the CD and HD schedulers converge are such greedy schedulers.

\subsection{Greedy Schedulers}\label{ssec:greedy}
The objective we consider is to maximise time-bounded reachability $Pr_{\S}^{\M}(l,t)$ for every location $l$ with respect to a particular scheduler class such as HD. 
Unfortunately, this optimisation problem is rather difficult to solve. 
Therefore, we start with analysing the special case of having little time left (that is, the remaining time $t$ is close~to~$0$). 
%

Time-abstract schedulers have no direct access to the time, but they can infer the distribution over the remaining time from the time-abstract history (or~its~length). 
When examining the resulting Poisson distribution one can easily see that for large step numbers the probability to take more than one further step declines faster than the probability to take exactly one further step. 
Thus, any increase of the likelihood of reaching the goal region sooner dominates the potential impact of reaching it in further steps (after sufficiently many steps). 

%
This motivates the introduction of greedy schedulers.
Schedulers are called greedy, if they (greedily) look for short-term gain, and favour it over any long-term effect.
Greedy schedulers that optimise the reachability within the first $k$ steps have been exploited in the efficient analysis of CTMDPs~\cite{Baier+all/05/efficientCTMDP}. 
To understand the principles of optimal control, however, a simpler form of greediness proves to be more appropriate: 
%
%
We call an HD scheduler \emph{greedy}
if it maximises the step probability vector of every location $l$ with respect to the lexicographic order (for example $(0,0.2,0.3,\dots)>_{\textit{lex}}(0,0.1,0.4,\dots)$).
To prove the existence of greedy schedulers, we draw from the fact that the supremum $d_l=\sup_{\S \in HD} d_{l,\S}$ obviously exists, where the supremum is to be read as a supremum with respect to the lexicographic order.
An action $a\in \act(l)$ is called \emph{greedy} for a location $l \notin B$ if it satisfies
$\shift(d_l)=\sum_{l'\in \locations}\probabilitymatrix(l,a,l')d_{l'}$, where $\shift(d_l)$ shifts the vector by one position (that is, $\shift(d_l)[i]=d_l[i+1]\;\; \forall i \in \mathbb N$).
For locations $l$ in the goal region $B$, all enabled actions $a\in \act(l)$ are greedy.

\begin{lemma}
Greedy schedulers exist, and they can be described as the class of schedulers that choose a greedy action upon every reachable time-abstract path.
\end{lemma}

\begin{proof}
It is plain that, for every non-goal location $l\notin B$, $\shift(d_l)\geq \sum_{l'\in \locations}\probabilitymatrix(l,a,l')d_{l'}$ holds for every action $a$, and that equality must hold for some.

For a scheduler $\S$ that always chooses greedy actions, a simple inductive argument shows that $d_l[i]=d_{l,\S}[i]$ holds for all $i \in \mathbb N$,
while it is easy to show that $d_l > d_{l,\S}$ holds if $\S$ deviates from greedy decisions upon a path that is possible under its own scheduling policy and does not contain a goal location.
\qed
\end{proof}

This allows in particular to fix a positional \emph{standard greedy scheduler} by fixing an arbitrary greedy action for every location.

To determine the set of greedy actions, let us consider a deterministic scheduler $\S$ that starts in a location $l$ with a non-greedy action $a$.
Then $\shift(d_{l,\S}) \leq \sum_{l'\in \locations}\probabilitymatrix(l,a,l') d_{l'}$ holds true, where
the sum $\sum_{l'\in \locations}\probabilitymatrix(l,a,l') d_{l'}$ corresponds to the scheduler choosing the non-greedy action $a$ at location $l$ and acting greedy in all further steps.
Let $d_{l,a}=\sum_{l'\in \locations}\probabilitymatrix(l,a,l') d_{l'}$ denote the step probability vector of such schedulers.

We know that $d_{l,\S}\leq d_{l,a}< d_l$. Hence, there is not only a difference between $d_{l,\S}$ and $d_l$, this difference will not occur at a higher index than the first difference between the newly defined $d_{l,a}$ and $d_l$.
The finite number of locations and actions thus implies the existence of a bound $k$ on the occurrence of this first difference between $d_{l,a}$ and $d_l$ as well as $d_{l,\S}$ and $d_l$.
While the existence of such a $k$ suffices to demonstrate the existence of optimal schedulers, we show in Subsection~\ref{ssec:constructing} that this constant $k<|\locations|$ is smaller than the CTMDP itself.

Having established such a bound $k$, it suffices to compare schedulers up to this bound. This provides us with the greedy actions, and also with the initial sequence $d_{l,a}[0], d_{l,a}[1],\dots, d_{l,a}[k]$ for all locations $l$ and actions $a$.
Consequently, we can determine a positive lower bound $\mu >0$ for the first non-zero entry of the vectors $d_l-d_{l,\S}$ (considering all non-greedy schedulers $\S$).
We call this lower bound $\mu$ the \emph{discriminator} of the CTMDP.
Intuitively, the discriminator $\mu$ represents the minimal advantage of the greedy strategy over non-greedy strategies.


\subsection{Uniform CTMDPs}\label{ssec:uniform}

In this subsection, we show that every CD or HD scheduler for a uniform CTMDP can be transformed into a scheduler that converges to 
this standard greedy scheduler.

In the quest for an optimal scheduler, it is useful to consider the fact that the maximal reachability probability can be computed using the step probability vector, because the likelihood that a particular number of steps happen in time $t$ is independent of the scheduler:
\begin{equation}
Pr_{\S}^{\M}(t) = \sum_{l\in \locations} \nu(l) \sum_{i=0}^\infty{d_{l,\S}[i]\cdot p_{\lambda t}(i)}.
\label{eq:uniformreachability}
\end{equation}

Moreover, the Poisson distribution $p_{\lambda t}$ has the useful property that the probability of taking $k$ steps is falling very fast. We define the \emph{greed bound} $\nM$ to be a natural number, for which
\begin{equation}
\mu \, p_{\lambda t}(n) \geq \sum_{i=1}^\infty p_{\lambda t}(n+i) 
\qquad \forall n\geq\nM
\label{eq:nM}
\end{equation}
holds true. It suffices to choose $\nM\geq\frac{2\lambda\, t}{\mu}$ since it implies $\mu p_{\lambda t}(n) \geq 2 p_{\lambda t}(n+1),\ \forall n>\nM$  (which yields (\ref{eq:nM}) by simple induction).
%
Such a greed bound implies that the decrease in likelihood of reaching the goal region in few steps caused by making a non-greedy decision after the greed bound dwarfs any potential later gain.
We use this observation to improve any given CD or HD scheduler $\S$ that makes a non-greedy decision~after ${\geq} \nM$ steps by replacing the behaviour after this history by a greedy scheduler.
Finally, we use the interchangeability of greedy schedulers to introduce a scheduler $\overline{\S}$ that makes the same decisions as $\S$ on short histories and follows the standard greedy scheduling policy once the length of the history reaches the greed bound.
For this scheduler, we show that $Pr_{\overline{\S}}^{\M}(t) {\geq} Pr_{\S}^{\M}(t)$ holds true.


\begin{theorem}
\label{thm:CDconvergence}
For uniform CTMDPs, there is an optimal scheduler for the classes CD and HD that converges to the standard greedy scheduler after $\nM$ steps.
\end{theorem}

\begin{proof}
Let us consider any HD scheduler $\S$ that makes a non-greedy decision after a time-abstract path $\pi$ of length $|\pi| \geq \nM$ with last location $l$.
If the path ends in, or has previously passed, the goal region, or if the probability of the history $\pi$ is $0$, that is, if it cannot occur with the scheduling policy of $\S$, then we can change the decision of $\S$ on every path starting with $\pi$ arbitrarily---and in particular to the standard greedy scheduler---without altering the reachability probability.

If $\Prob_{\S}^{\M}(\pi,t)>0$, then we change the decisions of the scheduler $\S$ for paths with prefix $\pi$ such that they comply with the standard greedy scheduler. We call the resulting HD scheduler $\S'$ and analyse the change in reachability probability using Equation~(\ref{eq:uniformreachability}):

\[
  \prob_{\S'}^{\M}(t) - \prob_{\S}^{\M}(t) = \Prob_{\S}^{\M}(\pi,t) \cdot \sum_{i=0}^\infty (d_l[i] -d_{l,\S_{\pi}}[i])\cdot p_{\lambda t}(|\pi|+i),
\]
where $\S_{\pi}: \pi' \mapsto \S(\pi \circ \pi')$ is the HD scheduler which prefixes its input with the path $\pi$ and then calls the scheduler $\S$.
The greedy criterion implies $d_l>d_{l,\S_{\pi}}$ with respect to the lexicographic order, and after rewriting the upper equation:
\[
 Pr_{\S'}^{\M}(t) - Pr_{\S}^{\M}(t) = \Prob_{\S}^{\M}(\pi,t) \cdot
   \left(\mu p_{\lambda t}(|\pi|+j) + \sum_{i>j}^\infty (d_l[i] -d_{l,\S_{\pi}}[i])\cdot p_{\lambda t}(|\pi|+i)\right)\quad \text{(for some $j>0$)}
\]
we can apply Equation~\ref{eq:nM} to deduce that the difference $Pr_{\S'}^{\M}(t) - Pr_{\S}^{\M}(t)$ is non-negative.

Likewise, we can concurrently change the scheduling policy to the standard greedy scheduler for all paths of length $\geq\nM$ for which the scheduler $\S$ makes non-greedy decisions. In this way, we obtain a scheduler $\S''$ that makes non-greedy decisions only in the first $\nM$ steps, and yields a (not necessarily strictly) better time-bounded reachability probability than $\S$.

Since all greedy schedulers are interchangeable without changing the time-bounded reachability probability (and even without altering the step probability vector), we can modify $\S''$ such that it follows the standard greedy scheduling policy after $\geq\nM$ steps, resulting in a scheduler $\overline{\S}$ that comes with the same time-bounded reachability probability as $\S''$.
Note that $\overline{\S}$ is counting if $\S$ is counting.

Hence, the supremum over the time-bounded reachability of all CD/HD schedulers is equivalent to the supremum over the bounded reachability of CD/HD schedulers that deviate from the standard greedy scheduler only in the first $\nM$ steps.
This class is finite, and the supremum over the bounded reachability is therefore the maximal bounded reachability obtained by one of its representatives.
\qed
\end{proof}

Hence, we have shown the existence of a---simple---optimal time-bounded CD scheduler. Using the fact that the suprema over the time-bounded reachability probability coincide for CD, CR, HD, and HR schedulers \cite{Baier+all/05/efficientCTMDP}, we can infer that such a scheduler is optimal for all of these classes. 

\begin{corollary}
\label{cor:CDCRHDHR}
$\max\limits_{\S \in CD} \prob_{\S}^{\M}(t) = 
\max\limits_{\S \in HR} \prob_{\S}^{\M}(t)$ holds for all uniform CTMDPs $\M$.
\qed
\end{corollary}


\subsection{Non-uniform CTMDPs}
\label{ssec:nonuniform}

Reasoning over non-uniform CTMDPs is harder than reasoning over uniform CTMDPs, because the likelihood of seeing exactly $k$ steps does not adhere to the simple Poisson distribution, but depends on the precise history. 
Even if two paths have the same length, they may imply different probability distributions over the time passed so far.
Knowing the time-abstract history therefore provides a scheduler with more information about the system's state than merely its length.
As a result, it is simple to construct example CTMDPs, for which history-dependent and counting schedulers can obtain different time-bounded reachability probabilities \cite{Baier+all/05/efficientCTMDP}.

In this subsection, we extend the results from the previous subsection to general CTMDPs.
We show that simple optimal CD/HD scheduler exist, and that randomisation does not yield an advantage:
\[
\max_{\S \in {CD}} \prob_{\S}^{\M}(t) = \max_{\S \in CR} \prob_{\S}^{\M}(t) \qquad \mbox{\text{and}} \qquad \max_{\S \in HD} \prob_{\S}^{\M}(t) = \max_{\S \in HR} \prob_{\S}^{\M}(t).
\]

To obtain this result, we work on the uniformisation $\U$ of $\M$ instead of working on $\M$ itself.
%
We argue that the behaviour of a general CTMDP $\M$ can be viewed as the observable behaviour of its uniformisation $\U$, using a scheduler that does not \emph{see} the new transitions and locations.
Schedulers from this class can then be replaced by (or viewed as) schedulers that do not \emph{use} the additional information.
And finally, we can approximate schedulers that do not use the additional information by schedulers that do not use it initially, where initially means until the number of visible steps---and hence in particular the number of steps---exceeds the greed bound $n_{\U}$ of the uniformisation $\U$ of~$\M$.
Comparable to the argument from the proof of Theorem \ref{thm:CDconvergence}, we show that  we can restrict our attention to the standard greedy scheduler after this initial phase, which leads again to a situation where considering a finite class of schedulers suffices to obtain the optimum.

\begin{lemma}
The greedy decisions and the step probability vector coincide for the observable and unobservable copy of each location in the uniformisation $\U$ of any CTMDP $\M$.
\end{lemma}

\begin{proof}
The observable and unobservable copy of each location reach the same successors under the same actions with the same transition rate.
\qed
\end{proof}

We can therefore choose a positional standard greedy scheduler whose decisions coincide for the observable and unobservable copy of each location.

For the \emph{uniformisation} $\U$ of a CTMDP $\M$,
we define the function $\vis: \pathsabs(\U) \rightarrow \pathsabs(\M)$ that maps a path $\pi$ of $\U$ to the corresponding path in $\M$, the \emph{visible path}, by deleting all unobservable locations and their directly preceding transitions from $\pi$.
(Note that all paths in $\U$ start in an observable location.)
We call a scheduler \emph{$n$-visible} if its decisions only depend on the visible path and coincide for the observable and unobservable copy of every location for all paths containing up to $n$ visible steps.
We call a scheduler \emph{visible} if it is $n$-visible for all $n\in \mathbb N$.

We call a HD/HR scheduler an ($n$-)visible HD/HR scheduler if it is ($n$-)visible, and we call an \mbox{($n$-)visible} HD/HR scheduler a visible CD/CR scheduler if its decisions depend only on the length of the visible path, and
an $n$-visible CD/CR scheduler if its decisions depend only on the length of the visible path for all paths containing up to $n$ visible steps.
The respective classes are denoted with according prefixes, for example, $n$-vCD.
Note that ($n$-)visible counting schedulers are not counting.

It is a simple observation that we can study visible CD, CR, HD, and HR schedulers on the uniformisation $\U$ of a CTMDP $\M$ instead of studying CD, CR, HD, and HR schedulers on $\M$.

\begin{lemma}
\label{lem:looping}
$\S \mapsto \S \circ \vis$ is a bijection from visible CD, CR, HD, or HR schedulers for the uniformisation $\U$ of a CTMDP $\M$ onto CD, CR, HD, or HR schedulers, respectively, of $\M$ that preserves the time-bounded reachability probability: $\prob_{\S}^{\U}(t) = \prob_{\S\circ \vis}^{\M}(t)$.
\qed
\end{lemma}

At the same time, copying the argument from the proof of Theorem \ref{thm:CDconvergence}, an $n_{\U}$-visible CD or HD scheduler $\S$ can be adjusted to the $n_{\U}$-visible CD or HD scheduler $\overline{\S}$ that deviates from $\S$ only in that it complies with the standard greedy scheduler for $\mathcal{U}$ after $n_{\U}$ visible steps, without decreasing the time-bounded reachability probability.
These schedulers are visible schedulers from a finite sub-class, and hence some representative of this class takes the optimal value.
We can, therefore, construct optimal CD and HD schedulers for every CTMDP $\M$.

\begin{lemma}
\label{lem:nlooping}
The following equations hold for the uniformisation $\U$ of a CTMDP $\M$:
\[
\max_{\S \in n_{\U}-\text{vCD}} Pr_{\S}^{\U}(t) = \max_{\S \in \text{vCD}} Pr_{\S}^{\U}(t)
 \quad \mbox{and} \quad
\max_{\S \in n_{\U}-\text{vHD}} Pr_{\S}^{\U}(t) = \max_{\S \in \text{vHD}} Pr_{\S}^{\U}(t).
\]
\end{lemma}

\begin{proof}
We have shown in Theorem \ref{thm:CDconvergence} that turning to the standard greedy scheduling policy after $n_{\U}$ or more steps can only increase the time-bounded reachability probability.
This implies 
that we can turn to the standard greedy scheduler after $n_{\U}$ \emph{visible} steps.

The scheduler resulting from this adjustment does not only remain $n_{\U}$-visible, it becomes a visible CD and HD scheduler, respectively.
Moreover, it is a scheduler from the finite subset of CD or HD schedulers, respectively, whose behaviour may only deviate from the standard scheduler within the first $n_{\U}$ visible steps.
\qed
\end{proof}

To prove that optimal CD and HD schedulers are also optimal CR and HR schedulers, respectively, we first prove the simpler lemma that this holds for $k$-bounded reachability.

\begin{lemma}
\label{lem:kbounded}
$k$-optimal CD or HD schedulers are also $k$-optimal CR or HR schedulers, respectively.
\end{lemma}

\begin{proof}
For a CTMDP $\M$ we can turn an arbitrary CR or HR scheduler $\S$ into a CD or HD scheduler $\S'$ with a time and $k$-bounded reachability probability that is at least as good as the one of $\S$ by first determinising the scheduler decisions from the \mbox{$(k+1)$st} step onwards---this has obviously no impact on $k$-bounded reachability---and then determinising the remaining randomised choices.

Replacing a single randomised decision on a path $\pi$ (for history-dependent schedulers) or on a set of paths $\Pi$ (for counting schedulers) that end(s) in a location $l$ is safe, because the time and $k$-bounded reachability probability of a scheduler is an affine combination---the affine combination defined by $\S(\pi)$ and $\S(|\pi|,l)$, respectively---of the $|\act(l)|$ schedulers resulting from determinising this single decision.
Hence, we can pick one of them whose time and $k$-bounded reachability probability is at least as high as the one of $\S$.

As the number of these randomised decisions is finite ($\leq k\,|\locations|$ for CR, and $\leq k^{|\locations|}$ for HR schedulers), this results in a deterministic scheduler after a finite number of improvement steps.
\qed
\end{proof}

\begin{theorem}
Optimal CD schedulers are also optimal CR schedulers.
\end{theorem}

\begin{proof}
First, for $n\rightarrow\infty$ the probability to reach the goal region $B$ in exactly $n$ or more than $n$ steps converges to $0$, independent of the scheduler.
Together with Lemma \ref{lem:kbounded}, this implies
\[\sup_{\S \in CR} Pr_{\S}^{\M}(t)=
 \lim_{n\rightarrow \infty} \sup_{\S \in CR} Pr_{\S}^{\M}(t;n) =
 \lim_{n\rightarrow \infty} \sup_{\S \in CD} Pr_{\S}^{\M}(t;n) \leq
 \max_{\S \in CD} Pr_{\S}^{\M}(t),\]
where equality is implied by $CD\subseteq CR$.
\qed
\end{proof}

Analogously, we can prove the similar theorem for history-dependent schedulers:
\begin{theorem}
Optimal HD schedulers are also optimal HR schedulers.
\qed
\end{theorem}

%
%

\subsection{Constructing Optimal Schedulers}\label{ssec:constructing}
The proof of the existence of an optimal scheduler is not constructive in two aspects. First, the computation of a positional greedy scheduler requires a bound for $k$, which indicated the maximal depth until which we have to compare the step probability vectors before we can ascertain equality. 
Second, we need an exact method to compare the quality of two (arbitrary) schedulers. 

\paragraph{A bound for $k$} 
The first property is captured in the following lemma. 
Without this lemma, we could only provide an algorithm that is guaranteed to converge to an optimal scheduler, but would be unable to determine whether an optimal solution has already been reached, as we never know when to stop when comparing step probability vectors.
In this lemma, however, we show that it suffices to check for equivalence of two step probability vectors only up to position $|\locations|-2$.
As discussed in Subsection~\ref{ssec:greedy}, this enables us to identify greedy actions and thus to \emph{compute} the discriminator $\mu$ and consequently the greed bound $\nM$.

\begin{lemma} 
Given a uniform CTMDP $\M$, the smallest $k$ that satisfies $\forall l\in\locations,\;a\in\act(l).\ d_l\neq d_{l,a} \Rightarrow \exists k'\leq k.\ d_l[k'] > d_{l,a}[k']$ is bounded by $|\locations| - 2$.
\end{lemma}
\begin{proof} 
The techniques we exploit in this proof draw from linear algebra, and are, while simple, a bit unusual in this context. We first turn to the simpler notion of Markov chains by resolving the nondeterminism in accordance with the positional standard greedy scheduler $\S$ whose existence was shown in Subsection~\ref{ssec:greedy}.

We first lift the step probability vector from locations to distributions, where $d_\nu=\sum_{l \in \locations}\nu(l)d_l$ is, for a distribution $\nu:\locations \rightarrow [0,1]$, the affine combination of the step probability vectors of the individual locations.
In this proof, we define two distributions $\nu,\nu':\locations \rightarrow [0,1]$ to be equivalent, if their step probability vectors $d_\nu=d_{\nu'}$ are equal.
Further, we call them $i$-step equivalent if they are equal up to position $i$ ($\forall j \leq i.\ d_\nu[j]=d_{\nu'}[j]$).

In order to argue with vector spaces, we extend these definitions to arbitrary vectors $\nu:\locations \rightarrow \mathbb R$ (instead of $\nu:\locations \rightarrow [0,1]$).

Let $D_i$ be the vector space spanned by $i$-step equivalent distributions $\nu,\nu'$ over $\locations$.
Naturally, $D_i \supseteq D_{i+1}$ always holds, as $i+1$ step equivalence implies $i$-step equivalence.
In addition we show that $D_0$ has $|\locations|-2$ dimensions, and that $D_i =  D_{i+i}$ implies that a fixed point is reached, which together implies that
$D_{|\locations|-2} = D_j$ for all $j \geq |\locations|-2$.
\begin{itemize}
 \item $D_0$ has $|\locations|-2$ dimensions:
$D_0$ is the vector space that contains the multitudes of differences $\delta=\lambda(\nu-\nu')$ of distributions $\nu,\nu':\locations \rightarrow [0,1]$ that are equally likely in the goal region (due to 0-step equivalency;  $d_\nu[0]=d_{\nu'}[0]$).

The fact that $\nu$ and $\nu'$ are distributions implies $\sum_{l \in \locations}\nu(l)=1$ and  $\sum_{l \in \locations}\nu'(l)=1$, and hence $\sum_{l \in \locations}\delta(l)=0$.
Further, the fact that $\nu$ and $\nu'$ are equally likely in the goal region implies $\sum_{l \in B}\nu(l)=\sum_{l \in B}\nu'(l)$, and hence $\sum_{l \in B}\delta(l)=0$.
Thus, $D_0$ has $|\locations|-2$ dimensions. (Assuming $B \neq \locations, B \neq \emptyset$, but otherwise every scheduler has equal quality.)

\item Once we have constructed $D_i$, we can construct the vector space $O_i$ that contains a vector $\delta$ if it is a multitude $\delta=\lambda(\nu-\nu')$ of differences $\nu-\nu'$ of distributions, such that $\shift(d_\nu)$ and $\shift(d_{\nu'})$ are $i$-step equivalent, that is, $\shift(d_\nu)-\shift(d_{\nu'})\in D_i$.

The transition from step probability vectors to the $\shift$ of them is a simple linear operation, which transforms the distributions according to the transition matrix of the embedded DTMC. 
Hence, we can obtain $O_i$ from $D_i$ by a simple linear transformation of the vector space.

\item Two step probability vectors are $i+1$-step equivalent if (1) they are $i$-step equivalent, and (2) their shift are $i$-step equivalent.
Therefore $D_{i+1} = D_i \cap O_i$ can be obtained by an intersection of the two vector spaces $D_i$ and $O_i$.
\end{itemize}

Naturally, this implies that the vector spaces are shrinking, that is, $D_0 \supseteq D_1 \supseteq \ldots \supseteq D_{|\locations|-2} \supseteq \ldots$, and that $D_i=D_{i+1}$ implies that a fixed point is reached. (It implies $O_i=O_{i+1}$ and hence $D_i=D_{j}\ (\forall j\geq i)$ by a simple inductive argument.) 

As $D_0$ is an $|\locations|-2$ dimensional vector space, and inequality ($D_i \neq D_{i+1}$) implies the loss of at least one dimension, a fixed point is reached after at most $|\locations|-2$ steps.
That is, two distributions are equivalent, if, and only if, they are $(|\locations|-2)$-step equivalent.
\\

Having established this, we apply it on the distribution $\nu_{l,a}$ obtained in one step from a position $l\notin B$ when choosing the action $a$, as compared to the distribution $\nu_l$ obtained when choosing the action according to the positional greedy scheduler.

Now, $d_l>d_{l,a}$ holds if, and only if $\shift(d_l)=d_{\nu_l}>d_{\nu_{l,a}}=\shift(d_{l,a})$, which implies $d_{\nu_l}[k']>d_{\nu_{l,a}}[k']$ for some $k'\leq |\locations|-2$, and hence $d_l[k]>d_{l,a}[k]$ for some $k < |\locations|$.
\qed
\end{proof}

\paragraph{Comparing schedulers} So far, we have narrowed down the set of candidates for the optimal scheduler to a finite number of schedulers. 
To determine the optimal scheduler, it now suffices to have a comparison method for their reachability probabilities. 

The combination of each of these schedulers with the respective CTMDP can be viewed as a \emph{finite} continuous-time Markov \emph{chain} (CTMC) since they behave like a positional scheduler after $\nM$ steps. 
Aziz et al.~\cite{Aziz/00/exactModelCheckingCTMC} have shown that the time-bounded reachability probability of CTMCs are computable (and comparable) finite sums $\sum_{i \in I} \eta_i e^{\delta_i}$, where the individual $\eta_i$ and $\delta_i$ are algebraic numbers.

We conclude with a constructive extension of our results:

\begin{corollary}
We can effectively construct optimal CD, CR, HD, and HR schedulers.
\hspace*{-4mm}\qed
\end{corollary}

\begin{corollary}
We can compute the time-bounded reachability probability of optimal schedulers as finite sums $\sum_{i \in I} \eta_i e^{\delta_i}$, where the $\eta_i$ and $\delta_i$ are algebraic numbers.
\qed
\end{corollary}

\subsubsection*{Complexity} These corollaries rely on the precise CTMC model checking approach of Aziz et al.~\cite{Aziz/00/exactModelCheckingCTMC}, which only demonstrates the effective decidability of this problem.
We deem it unlikely that a complexity for finding optimal strategies can be provided prior to determining the respective CTMC model checking complexity.

	\definecolor{darkergreen}{rgb}{0,.6,0}

\subsection{Example}
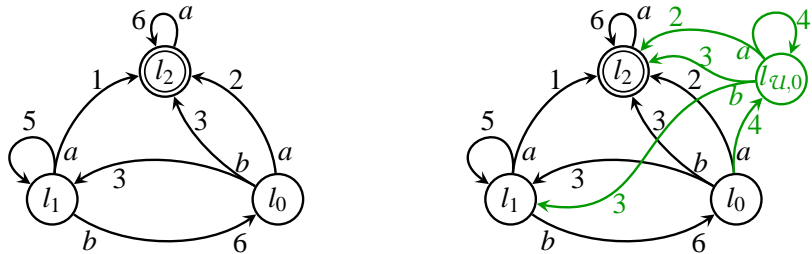
\begin{figure}[t]
\newcommand{\automaton}{
	\psset{arrowsize=5pt,nodesep=0pt,arrowlength=1,linewidth=.9pt}
    
    \cnode[](4.5,1.1){10pt}{0}
    \rput(4.5,1.1){$l_0$}
    
    \cnode[](1.5,1.1){10pt}{1}
    \rput(1.5,1.1){$l_1$}
    
    \cnode[](3,2.8){10pt}{2}
    \cnode[linewidth=.65pt](3,2.8){8pt}{inner2} 
    \rput(3,2.8){$l_2$}
    
    \nccurve[angleA=150,angleB=35]{->}{0}{1}
    \nccurve[angleA=150,angleB=-70]{->}{0}{2}
    
    \nccurve[angleA=95,angleB=-10]{->}{0}{2}
    
    \nccurve[angleA=85,angleB=-170]{->}{1}{2}
    \nccurve[angleA=85,angleB=145,ncurv=5]{->}{1}{1}

    \nccurve[angleA=-35,angleB=-145]{->}{1}{0}
    
    \nccurve[angleA=70,angleB=110,ncurv=5]{->}{2}{2}

    
    \rput(4.6,1.7){$a$} 
    \rput(4.03,1.6){$b$} 
    \rput(2,.55){$b$} 
    \rput(1.75,1.7){$a$} 
	\rput(3.28,3.58){$a$} 
    
    \rput(3.95,2.7){$2$} 
    \rput(3.48,2.15){$3$} 
    \rput(2.4,1.38){$3$} 
    \rput(4,.53){$6$} 
    \rput(2.1,2.7){$1$} 
    \rput(1.2,2.15){$5$} 
    \rput(2.65,3.5){$6$} 
}
\centering
\begin{pspicture}(5.6,3.0)
	\rput(-.6,-.7){
		\automaton
	}
\end{pspicture}
\begin{pspicture}(5.6,3.0)
	\rput(-.2,-.7){
	\automaton
	\psset{linecolor=darkergreen}
	\cnode[linewidth=.9pt](5.1,2.7){10pt}{0U}
    \rput(5.1,2.7){\color{darkergreen}$l_{\U,0}$}
    
    \nccurve[angleA=95,angleB=-135,ncurvB=.8]{->}{0}{0U}
	\rput(4.75,2.1){\color{darkergreen}$4$}
	
	\nccurve[angleA=130,angleB=50]{->}{0U}{2}
	\nccurve[angleA=130,angleB=70,ncurv=5]{->}{0U}{0U}
	\rput(5.4,3.5){\color{darkergreen}$4$}
	\rput(3.7,3.53){\color{darkergreen}$2$}
	\rput(4.6,3.05){\color{darkergreen}$a$} 
	
	\nccurve[angleA=-175,angleB=20,ncurvA=1,ncurvB=1]{->}{0U}{2}
	\nccurve[angleA=-175,angleB=-10,ncurvA=1,ncurvB=1]{->}{0U}{1}
	\rput(4.1,3){\color{darkergreen}$3$}
	\rput(2.95,1.05){\color{darkergreen}$3$}
	\rput(4.5,2.48){\color{darkergreen}$b$} 
	
	}
\end{pspicture}
\caption{The example CTMDP $\mathcal M$ (left) and the reachable part of its uniformisation $\mathcal U$ (right).}
\label{fig:example}
\end{figure}

To exemplify our proposed construction, let us consider the example CTMDP $\M$ depicted in Figure~\ref{fig:example}.
As $\M$ is not uniform, we start with constructing the uniformisation $\U$ of $\M$  (cf.\ Figure~\ref{fig:example}).

$\U$ has the uniform transition rate $\lambda=6$.
Independent of the initial distribution of $\M$, the unobservable copies of $l_1$ and $l_2$ are not reachable in $\U$, because the initial distribution of a uniformisation assigns all probability weight to observable locations, and the transition rate of all enabled actions in $l_1$ and $l_2$ in $\M$ is already $\lambda$. (Unobservable copies of a location $l$ are only reachable from the observable and unobservable copy of $l$ upon enabled actions $a$ with non-maximal exit rate $\ratematrix(l,a,\locations) \neq\lambda$.)

Disregarding the unreachable part of $\U$, there are only $8$ positional schedulers for $\U$, and only $4$ of them are visible (that is, coincide on $l_0$ and {\color{darkergreen}$l_{\U,0}$}).
They can be characterised by $\S_1=\set{l_0\mapsto a,\ l_1\mapsto a}$, $\S_2=\set{l_0\mapsto a,\ l_1\mapsto b}$, $\S_3=\set{l_0\mapsto b,\ l_1\mapsto a}$, and  $\S_4=\set{l_0\mapsto b,\ l_1\mapsto b}$.
In order to determine a greedy scheduler, we first determine step probability vectors: 

For $l_0$: $d_{l_0,\S_1}=d_{l_0,\S_2}=(\frac{1}{3},\frac{5}{9},\frac{19}{27},\dots)$, 
$d_{l_0,\S_3}=(\frac{1}{2},\frac{7}{12},\frac{43}{72},\dots)$, 
$d_{l_0,\S_4}=(\frac{1}{2},\frac{1}{2},\frac{3}{4},\dots)$. 

For $l_1$: $d_{l_1,\S_1}=d_{l_1,\S_3}=(\frac{1}{6},\frac{7}{36},\frac{71}{216},\dots)$, 
$d_{l_1,\S_2}=(0,\frac{1}{3},\frac{5}{9},\dots)$, 
$d_{l_1,\S_4}=(0,\frac{1}{2},\frac{1}{2},\dots)$. 

Note that, in the given example, it suffices to compute the step probability vector for a single step to determine that $\S_3$ is optimal (w.r.t.\ the greedy optimality criterion); in general, it suffices to consider as many steps as the CTMDP has locations.
Since deviating from $S_3$ decreases the chance to reach the goal location $l_2$ in a single step by $\frac{1}{6}$ both from $l_0$ and $l_1$, the discriminator $\mu=\frac{1}{6}$ is easy to compute.

Our coarse estimation provides a greed bound of $n_\U=\lceil72\cdot t\rceil$, where $t$ is the time bound, but $n_\U=\lceil 42\cdot t\rceil$ suffices to satisfy Equation~(\ref{eq:nM}).

When seeking optimal schedulers from any of the discussed classes, we can focus on the finite set of those schedulers that comply with $\S_3$ after $n_\U$ (visible) steps.
In the previous subsection, we describe how the precise model checking technique of Aziz et al.~\cite{Aziz/00/exactModelCheckingCTMC} can be exploited to turn the existence proof into an effective technique for the construction of optimal schedulers.

\section{Extension to Continuous-Time Markov Games}\label{sect:games}
Markov decision processes can easily be extended to continuous-time Markov games (CTGs) $\G=(\locations_A,\locations_D,\act,\ratematrix,\nu,B)$ by disintegrating the set of  locations into game positions of a maximiser ($\locations_A$, angelic game positions) and a minimiser ($\locations_D$, demonic game positions). 
These two players have antagonistic objectives to maximise and minimise the time-bounded reachability probability. 
These games are closely related to the CTMDP framework, and we define, for a given Markov game $\G$, the \emph{underlying} CTMDP $\M=(\locations_A\dot\cup\locations_D,\act,\ratematrix,\nu,B)$.
CTGs are called \emph{uniform} if their underlying CTMDP is uniform.

The players can choose an action upon the entrance to one of their locations, and, as with schedulers for CTMDPs, they may have limited access to the timed history of the system. 
We only consider time-abstract strategies $S_X:\pathsabs^X(\G)\to\dist(\act)$ for both players, where paths are defined over the underlying CTMDP, and $\pathsabs^X(\G)$ (for $X\in\set{A,D}$) is the set of paths that end with a location in $\locations_X$.

Obviously, there is a one-to-one mapping between \emph{combined strategies} 
\[\S_{A+D}(\pi)=\left\{\begin{array}{l} \S_A(\pi)\qquad \text{ if }\pi\in\pathsabs^A(\G) \\ \S_D(\pi)\qquad \text{ if }\pi\in\pathsabs^D(\G) \end{array} \right.\] 
of a CTG and schedulers of the underlying CTMDP.

For a given CTG and a pair of strategies $\S_A$, $\S_D$ we define the according probability space equivalent to the probability space of the underlying CTMDP with the combined strategy $S_{A+D}$. Then, the time-bounded reachability probability can be formulated for CTGs as follows: 
\begin{equation}
\sup_{\S_A} \inf_{\S_D} \prob_{\S_{A+D}}^{\G}(t) = \inf_{\S_D} \sup_{\S_A} \prob_{\S_{A+D}}^{\G}(t)
\label{eq:supinfsup}
\end{equation}
where equality is guaranteed by~\cite[Theorem 3.1]{brazdil_et_al:LIPIcs:2009:2307}.
%

For uniform CTGs, a theorem similar to Theorem~\ref{thm:CDconvergence} has recently been shown:

\begin{theorem}
\label{thm:unigames}
\cite{brazdil_et_al:LIPIcs:2009:2307}
For uniform CTGs $\G$ with counting strategies, we can compute a bound $n_{\G}$ (comparable to our greed bound) and a memoryless deterministic greedy strategy $\S:\locations\to\act$, such that following $\S$ is optimal for both players after $n_{\G}$ steps.
\end{theorem}

That is, optimal (counting) strategies for uniform Markov games have a similarly simple structure as those for CTMDPs.
Now, we extend these results to history-dependent (HD and HR) schedulers:

\begin{theorem} The optimal CD strategies from Theorem~\ref{thm:unigames} (that is, for uniform CTGs) are also optimal HR strategies. 
\end{theorem}

\begin{proof}
Let us assume the minimiser plays in accordance with her optimal CD strategy.
Let us further assume that the maximiser has an HR strategy that yields a better result than his CD strategy.
Then it must improve over his optimal CD strategy by a margin of some $\varepsilon$.

Let us define $p(k,l)$ as the maximum of the probabilities to still reach the goal region in the future that the maximiser can reach under the paths of length $k$ which end in location $l$ with the \emph{better} history dependent strategy. Further, let $h_l(k)$ be a path where this optimal value is taken. (Note that our goal region is absorbing.)
The decision this HR scheduler takes is an affine combination of deterministic decisions, and the quality (the probability of reaching the goal region in the future) is the respective affine combination of the outcome of these pure decisions. Hence, there is at least one pure decision that (not necessarily strictly) improves over the randomised decision. 

As our CTG is uniform, we can improve this history dependent scheduler by changing all decisions it makes on a path $\pi=\pi'_l\circ\pi'$ that start with a path $\pi'_l$ of length 2 ending in a location $l$, to the decisions it made upon the path $h_l(2)\circ\pi'$.
(The improvement is not necessarily strict.) 
We then improve it further (again not necessarily strictly) by turning to the improved pure decision. The resulting strategy is initially counting---it depends only on the length of the history and the current location---and deterministic for paths up to length 2.

Having constructed a history dependent scheduler that is initially counting and deterministic for paths up to length $k$, we repeat this step for paths $\pi=\pi'_l \circ\pi'$ that start with a history $\pi'_l$ of length $k+1$, where we replace the decision made by our initially $k$ counting and deterministic scheduler by the decision made on  $h_l(k+1)\circ\pi'$, and then further to its deterministic improvement.
This again leads to a---not necessarily strict---improvement.

Once the probability of making at least $k$ steps falls below $\varepsilon$, any deterministic counting scheduler that agrees on the first $k$ steps with a history dependent scheduler from this sequence (which is initially counting and deterministic for at least $k$ steps) improves over the counting scheduler we started with for the maximiser, which contradicts its optimality.

A similar argument can be made for the minimiser. \qed
\end{proof}


Our argument that infers the existence of optimal strategies for general CTMDPs from the existence of optimal strategies for uniform CTMDPs does not depend on the fact that we have only one player with a particular objective.
In fact, it can be lifted easily to Markov games.

\begin{theorem}
For a Markov game $\mathcal G$, we can effectively construct optimal CD and HD schedulers, which are also optimal CR and HR schedulers, respectively, and
we can compute the time-bounded reachability probability of optimal schedulers as finite sums $\sum_{i \in I} \eta_i e^{\delta_i}$, where the $\eta_i$ and $\delta_i$ are algebraic numbers.
\end{theorem}

\begin{proofsketch}
We start again with the uniformisation $\U$ of the Markov game $\G$. By Theorem~\ref{thm:unigames}, there is a deterministic memoryless greedy strategy for both players in $\U$ that is optimal after $n_\U$ steps.
Hence, we can argue along the same lines as for CTMDPs:
\begin{itemize}
\item We study the \emph{visible} strategies on the uniformisation $\U$ of $\G$. 
Like in the constructions from Section~\ref{ssec:nonuniform}, we use a bijection $\vis$ from the visible strategies on $\U$ onto the strategies of $\G$, which preserves the time-bounded reachability.

\item We define $n_\U$-visible strategies analogously to the $n_\U$-visible schedulers to be those strategies, which can use the additional information provided by $\U$ after $n_\U$ visible steps have passed.

\hspace{0.025\textwidth} After $n_\U$ visible steps, the class of $n_\U$-visible strategies clearly contains the deterministic greedy strategies described in the previous theorems of this section, as they can use all information after step $n_\U$. 
Using Theorem~\ref{thm:unigames} we can deduce that, for both players, it suffices to seek an optimal $n_\U$-visible strategy in the subset of those strategies that turn to the \emph{standard greedy strategy} after $n_\U$ \emph{visible} steps.


\item Locations $l$ and their counterparts $l_\U$ have exactly the same exit rates for all actions, and therefore a greedy-optimal memoryless strategy will pick the same action for both locations (up to equal quality of actions). This directly implies that the standard greedy scheduler is a visible strategy, and with it all $n_\U$-visible strategies that turn to the standard greedy strategy after $n_\U$ visible steps are visible strategies.
Hence, an optimal strategy for the class of $n_\U$-visible strategies that turn to the standard greedy strategy after $n_\U$ visible steps is also optimal for the class of visible strategies (time-abstract strategies in $\G$, respectively). 

\item 
For deterministic strategies, this class is finite, which immediately implies the existence of an optimum in this class (using Equation~\ref{eq:supinfsup}).
\end{itemize}

Randomised strategies again cannot provide an advantage over deterministic ones, because their outcome is just an affine combination of the outcome of the respective pure strategies, and the extreme points are taken
at the fringe. (Technically, we can start with any randomised strategy and replace one randomised decision after another by a pure counterpart, improving the quality of the outcome---not necessarily strictly---for the respective player.)

Consequently, we are left with a finite set of history dependent or counting candidate
strategies, respectively, and the result can---at least in principle---be found by applying a brute force
approach: For each of these deterministic strategies, we can compute the reachability probability using the algorithm of Aziz et al.~\cite{Aziz/00/exactModelCheckingCTMC}, which allows for identifying the deterministic strategies that mark an optimal Nash equilibrium. \qed

\end{proofsketch}

%
%
%

\bibliographystyle{plain}
\bibliography{bib}

\end{document}